\documentclass[review]{elsarticle}
\usepackage[utf8]{inputenc}

\usepackage{hyperref}
\usepackage{subfig}
\journal{Safety Science}

\begin{document}

\begin{frontmatter}

\title{Impact of baggage collection behaviour on \\aircraft evacuation}

\author[1]{Dan Hodgson}
\author[2]{Christian Tonge}
\author[1]{Martyn Amos\corref{cor1}}

\address[1]{Northumbria University, Newcastle upon Tyne, UK}
\address[2]{Arup, Manchester, UK.}
\cortext[cor1]{Corresponding author: martyn.amos@northumbria.ac.uk}

\begin{abstract}
Recent reports of emergency aircraft evacuations have highlighted an increasing tendency amongst evacuees to ignore clear safety warnings and to collect and carry personal items of baggage during egress. However, relatively little work has so far been done on quantifying the impact of such behaviour on the evacuation process. In this paper, we report the results of validated simulation experiments (using the Boeing 777 wide-body aircraft), which confirm that even a relatively low level of baggage collection can significantly delay evacuation. Our platform provides one possible framework for the investigation of processes and mitigation tactics to minimise the impact of baggage collection behaviour in future. 
\end{abstract}
\end{frontmatter}

\section{Introduction}

In emergency situations where occupants are uninjured and the cabin environment is intact, survival of passengers is largely determined by their ability to move from seat to exit, subject to the time limits imposed by environmental factors such as fire and smoke \cite{snow1970survival}. Other non-environmental aspects of evacuation that influence survival include {\it configurational} factors (e.g., the internal layout of the cabin and how it affects passenger flow and access to exits), {\it procedural} factors (e.g., experience and training of cabin crew, and their ability to enact protocols), and {\it biobehavioural} factors (e.g., the physical attributes and behavioural responses of evacuees) \cite{snow1970survival}. In this paper, we focus on behavioural aspects; specifically, we study the impact of {\it baggage collection} during evacuation.

Baggage collection can severely impact the effectiveness and speed of execution of evacuation protocols. Events such as the British Airways Boeing 777 fire at Las Vegas (September 8, 2015), the Emirates Boeing 777 crash at Dubai (August 3, 2016, in which one firefighter died during the rescue operation) and the Aeroflot Sukhoi Superjet 100 crash in Russia (May 5, 2019, in which 41 people died),  have all focused attention on the problem of passengers collecting items of luggage prior to and during the evacuation process. In each of the three cases cited (and in many other documented cases), witnesses stated that a significant number of passengers ignored clear safety instructions to leave behind any personal belongings.

Despite being acknowledged as a clear and significant issue in aircraft evacuation, surprisingly few studies have investigated the impact of baggage collection, and we are unaware of any relevant studies that specifically consider wide-body (dual-aisle) aircraft. In this paper, we describe a computational simulation of the evacuation of a Boeing 777 (validated against the certification trial) which incorporates baggage collection. Our findings confirm that, for a realistically populated aircraft, even a very modest frequency of baggage collection can cause the total evacuation time to significantly exceed the certification threshold. Our platform therefore provides a framework for future investigations into strategies for the assessment, mitigation and/or prevention of baggage collection during emergency aircraft evacuations.

The rest of the paper is organized as follows: in Section ~\ref{background} we discuss the issue of baggage collection in the context of emergency aircraft evacuation, and discuss existing work in this area. This motivates the construction of a computational simulation framework to study the impact of baggage collection behaviour, which is described in Section ~\ref{methods}. We describe our results in Section ~\ref{results}, and conclude in Section ~\ref{conclusion} with a discussion of the implications of our results and some suggestions for future work.

\section{Background and motivation}
\label{background}

All airlines advise passengers not to take personal belongings during an aircraft evacuation. This message is emphasised during in-flight briefings, on seat information cards, and during an evacuation process itself. However, this message does not prevent a significant number of individuals from persisting in collecting personal items during evacuations. A survey by the US National Transportation Safety Board (NTSB) of cabin crew and passengers involved in 46 different evacuation events found that ``Passengers exiting with carry-on baggage were the most frequently cited obstruction to evacuation", and nearly 50$\%$ of passengers with carry-on luggage admitted to having attempted to retrieve a bag during their evacuation \cite{NTSB2000}. Significant delays and obstruction were caused by episodes such as arguments between cabin crew and passengers over luggage retrieval, the blocking of aisles while items were retrieved from overhead lockers, and the manoeuvring of objects through emergency exits (in one case, an evacuation was delayed by an individual attempting to carry a garment bag through an exit). Objects can also themselves block aisles, limit the movement of the person carrying them, damage emergency slides, and serve as projectiles, causing injury to the carrier or other passengers.

The Royal Aeronautical Society (RAS) highlighted this problem in a 2018 report \cite{RAS2018}, but also pointed out that ``Such actions by passengers are not new" (citing an incident dating back to 1984). As argued in the report, the {\it perception} of an increase in baggage collection behaviour may be due to the contemporary ubiquity of mobile phones and social media, with footage of incidents being shared rapidly and widely. Clearly, though, there is a growing trend for baggage collection behaviour in aircraft evacuation, and the RAS offer several reasons why this may be the case. These may be summarised as follows: some aircraft manufacturers are increasing the size and capacity of in-cabin overhead lockers, which means that some items that would previously have been inaccessible in the hold during an evacuation are now available to take. Many carriers now charge a fee to carry items in the hold (driven by the need to reduce turn-round times), and the rise in availability of low-cost, short-haul flights means that cabin bags are now effectively the only choice of luggage for a much higher proportion of total trips. As a result, the carry-on bag is often the only item in a passenger's possession (which may explain their reluctance to leave it behind, especially if it contains travel documents, medication, etc.) Additionally, the ubiquity of high-value items such as tablet and laptop computers, mobile phones and cameras means that individuals are much more likely to want to keep such items close to their person. Finally, many airports have significantly expanded their airside commercial operations in recent times, with the result that passengers are buying products after check-in, further contributing to problems with cabin stowage.

Both reports we cite here \cite{NTSB2000,RAS2018} supply a lengthy catalogue of incidents in which the collection of personal items has significantly affected the emergency evacuation of an aircraft. In 2000, the NTSB argued that ``...passengers who attempt to take their luggage during evacuations continue to present undue risks and delays to a successful evacuation. By retrieving luggage during an evacuation, passengers increase the potential for serious injuries or loss of life. The Safety Board concludes that passengers’ efforts to evacuate an airplane with their carry-on baggage continue to pose a problem for flight attendants and are a serious risk to a successful evacuation of an airplane." More recently, the RAS \cite{RAS2018} observed that ``It would appear that cabin crew have little control over passengers who insist on taking cabin baggage with them in an evacuation. Perhaps it is only a matter of time before an evacuation occurs when the issue of cabin baggage becomes a survival factor."

 The current paper was motivated by a specific incident that became relatively notorious in the popular press, due to the behaviour of passengers during the evacuation. On September 8th 2015, a British Airways Boeing 777-200 was preparing to leave Las Vegas McCarran International Airport, with 157 passengers and 17 crew on board. During takeoff, the aircraft suffered a failure of its left engine, which led to a fire. The crew aborted the takeoff, stopped the aircraft, and evacuated the aircraft. Although there were no fatalities, one crew member received serious injuries \cite{NTSBVegas}.

\begin{figure}[!htb]
\begin{centering}
\includegraphics[width=2.5in]{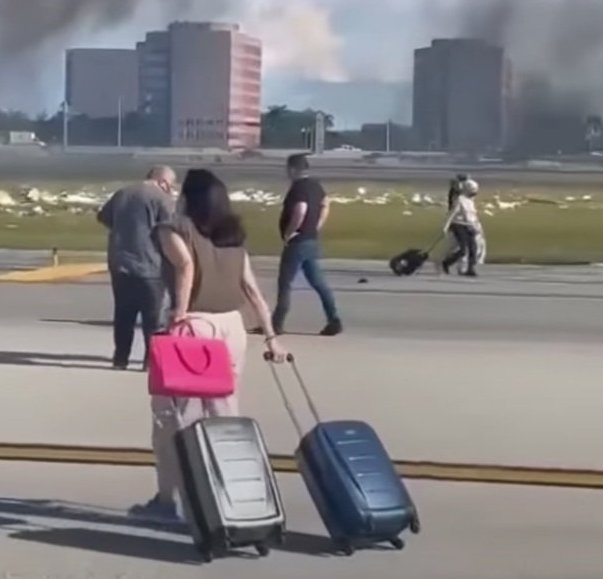}
\caption{Screengrab taken from footage of the RED Air Flight 203 incident, which shows passengers with large carry-on bags post-evacuation.}
\label{fig:extreme}
\end{centering}
\end{figure}

Although a significant event in its own right, the incident received particular attention, due to the number of passengers who were observed carrying luggage during and immediately after the evacuation. According to one BBC report, ``People streaming out of the plane were holding purses, flip-flop sandals, rolling bags and shoulder bags", and a significant backlash ensued on social media and in the press. The official NTSB report \cite{NTSBVegas} did not highlight baggage collection as a specific concern, stating that ``The flight attendants at the two most-used exits ... recalled seeing very little baggage at their exits." However, video footage taken just after the event does appear to show a significant proportion (on the order of 30$\%$) of passengers carrying personal items and/or luggage, and the report stated that the NTSB ``...remains concerned about the safety issues resulting from passengers evacuating with carry-on baggage, which could potentially slow the egress of passengers and block an exit during an emergency." The report also highlighted the fact that the aircraft was only half-full. After this episode, a number of related incidents occurred around the world during which passengers stopped to collect luggage during the evacuation process. Specific examples of note include Emirates Flight EK521 (August 3, 2016, Dubai, UAE \cite{wells}), Aeroflot Flight 1492 (May 5, 2019, Sheremetyevo, Russia, 41/78 pax killed \cite{kaminski}), and RED Air Flight 203 (June 21, 2022, Miami, USA \cite{thornber}) (see Figure ~\ref{fig:extreme} for a screengrab of a movie{\footnote{\url{https://twitter.com/rawsalerts/status/1539391173781073922}}} taken during this incident; this graphically illustrates an extreme example of the type of behaviour to which we refer in this paper). 

On March 1, 2019, a Lauda Air Airbus A320-214 was subject to an aborted take-off from London Stansted Airport, and the aircraft was evacuated. In its report \cite{lauda} on the incident, the UK's Air Accidents Investigation Branch highlighted that ``Numerous passengers also commented that the aisle and Doors 2L and 2R and the overwing exits, were impeded as people were trying to take their baggage from under seats and overhead bins. As a result, passengers were shouted at by some to leave their baggage behind. One passenger thought that about half of the passengers took their hand baggage with them. Images of passengers leaving the aircraft with baggage from the right overwing exit were captured by the ... onboard infrared CCTV camera."

\subsection{Previous work}

Given its clear significance, the issue of baggage collection during emergency aircraft evacuation presents a clear opportunity for research. However, the phenomenon has, to date, received relatively little attention in the scientific literature. Consideration of passengers and their luggage in the context of aircraft has generally centred on its impact on the speed and efficiency of the {\it boarding} process \cite{best2014ped,tang2018aircraft}. Where luggage is considered in connection with aircraft emergencies, it is generally in the context of {\it training} passengers \cite{chittaro2014desktop}, although some recent work \cite{giitsidis2017parallel,johansson2019modelling,song2023emergency} has considered its impact on evacuation. None of this work was conducted on models of wide-body aircraft (we exclude \cite{lee2021effects} from consideration, as there are significant gaps in its description of the methodology).

In this paper, we study the impact of baggage collection behaviours in the context of the evacuation of wide-body aircraft, which is the first such rigourous study. We first evaluate a baseline simulation of a Boeing 777 against the only documented certification trial, before showing how this may be used to investigate the effect of varying levels of aircraft occupancy and baggage collection. We conclude with a discussion of our results, and consider their implications in the wider context of commercial aviation.

\section{Methods}
\label{methods}

In the experiments that follow, we simulate the evacuation of a Boeing 777-200 aircraft. The ``Triple Seven" is a long-range, wide-body, twin-engine commercial airliner, which first saw service in 1995. Produced in a number of variants, the 777 model remains a popular choice for airlines, and over 1500 have been delivered to date. 

National aviation bodies require all commercial aircraft to receive certification before carrying passengers. As part of the certification process, an {\it evacuation trial} must be performed, in order to demonstrate the feasibility of evacuating a full complement of passengers and crew within 90 seconds, with only half of the aircraft's exits available (in order to simulate a fire or other impediment). 
The results of the first 777-200 evacuation trial (which was performed in February 1995) are discussed in depth in \cite{RAS2018} (as well as in an informal paper by a representative of Boeing \cite{Satterfield}); this reports that the final evacuee touched the ground after 94.3 seconds, and 419 of the 420 ``passengers" left within 85.7 seconds. Two subsequent trials were performed, but we have limited information available on important aspects of those events, such as the specific exits that were available. For those reasons, we use \cite{RAS2018,Satterfield} as the basis for comparison.

In what follows, we use the Pathfinder software package{\footnote{https://www.thunderheadeng.com/pathfinder/}}, produced by Thunderhead Engineering, which is a well-established platform for agent-based simulation of the evacuation of buildings, stadia, aircraft, and other spaces \cite{cuesta2015collection,ronchi2016modelling,wang2014emergency}. The package uses occupant ``profiles", which store physical characteristics of agents, and ``behaviours", which allow us to assign objectives and probabilistic rule-following actions to individuals. 

\subsection{Aircraft configuration}

\begin{figure}[!htb]
\begin{centering}
\includegraphics[width=\textwidth]{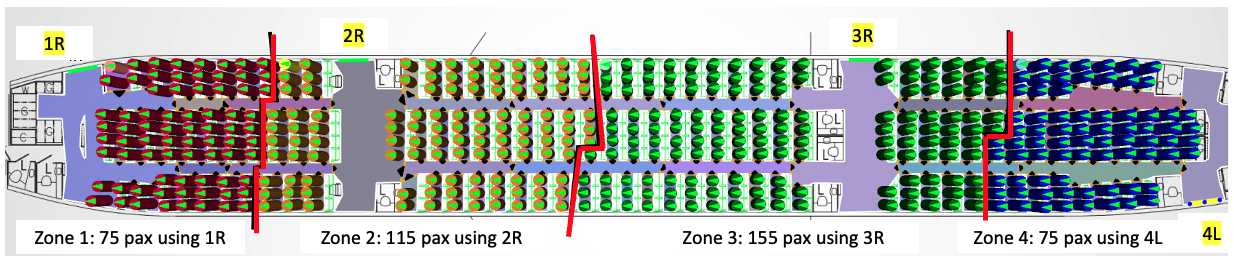}
\caption{Aircraft layout for certification trial experiments (fore-aft shown left-right). Available exits (1R, 2R, 3R and 4L) are highlighted and used to construct four internal zones for the purposes of exit allocation to individual passengers.}
\label{fig:cert_layout}
\end{centering}
\end{figure}

We first consider the basic internal configuration of the aircraft; it is generally the case that certification trial layouts aim to maximise the number of passengers on board (in order to obtain the largest possible upper bound). In Figure ~\ref{fig:cert_layout} we show the aircraft layout, as constructed in Pathfinder. The actual certification trial layout is not specified in either report \cite{RAS2018,Satterfield}, so we modified a known 440-seat configuration from \cite{Boeing777} to produce a plausible 420-seat configuration. In what follows, as in the certification trial, the only accessible doors are 1R, 2R, 3R and 4L (i.e., the first three starboard exits, and the rear port exit). We divide the aircraft into four ``zones", each zone corresponding to an exit, and passengers in each zone are instructed to move towards that exit when the evacuation commences; Zone 1 (1R, at the front of the aircraft) has 75 passengers (pax), Zone 2 (2R, front-middle) has 115 pax, Zone 3 (3R, rear-middle) has 155 pax, and Zone 4 (rear, 4L) has 75 pax. Most allocations were made on the basis of spatial proximity alone, but a small number of allocations were made on the basis of both spatial proximity and a need to avoid over-loading certain exits (so, for example, passengers in row 8 are allocated to 1R on the basis of load balancing). However, this latter case accounts for a relatively small number of passengers.

\begin{figure}[!htb]
\begin{centering}
\includegraphics[width=2in]{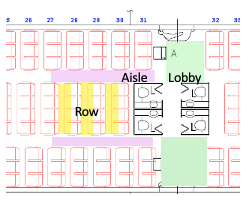}
\caption{Section of aircraft layout, showing different region types.}
\label{fig:spaces}
\end{centering}
\end{figure}

\newpage
The aircraft representation was created as follows; first, a graphical schematic was imported into Pathfinder, and this was used to guide the placement of functional regions (such as seats, aisles and exits) within the software's representation scheme, which were connected by virtual ``doors" (these are different to the aircraft exits). Although it is possible to model the entire cabin space as a single ``room", we found (during prototyping) that this caused issues with the decision-making processes of individual agents. Specifically, using a single large space prevented agents from considering alternative routes when their preferred exit was congested. The layout of the aircraft was divided into three different types of {\it region}; (1) {\it aisle} (self-explanatory), (2) {\it lobby} (large open spaces, generally in the vicinity of exits, but also around galleys and toilets), and (3) {\it row} (spaces between seats) (Figure ~\ref{fig:spaces}). Each of these region types affects the speed at which passengers move, as we describe shortly.

We derive all aircraft data from the official Boeing guidance \cite{Boeing777}.
In Table ~\ref{tab:size} we specify the dimensions of various internal features of the aircraft that are pertinent to our simulation. We note, in particular, that the 777-200 has uniform exit door widths, and there are no ``over-wing" exits. Due to certain limitations of the software, the seat row width (that is, the distance between two rows of seats) was defined as being equal to the width of one average passenger. 

\begin{table}[]
\begin{centering}
\begin{tabular}{|l|l|}
\hline
\textbf{Feature} & \textbf{Size} \\
\hline
Cabin width      & 5.84m         \\ \hline
Seat width       & 0.43m         \\ \hline
Seat row space width   & 0.33m         \\ \hline
Aisle width      & 0.43m         \\ \hline
Exit door width  & 1.23m         \\ \hline
\end{tabular}
\caption{Internal feature dimensions.}
\label{tab:size}
\end{centering}
\end{table}

\subsection{Baseline passenger behaviour}

We use the Pathfinder terminology of ``profiles" to describe passenger attribute sets, and ``behaviours" to describe their goal-seeking actions.  Passengers are modelled as cylinders, which is the default option. We use a passenger diameter value of 45.58cm, which is in line with the reported shoulder width (45.4cm) observed in the experiments described in \cite{melis2020effect}. Pathfinder agents may temporarily reduce their diameter in order to resolve congestion (essentially, agents may attempt to ``squeeze past" one another), and we enable this option, with a minimum diameter of 33cm set.

The default {\it nobags} profile is described in Table ~\ref{tab:nobags}. We apply space-specific modifiers (i.e., multipliers) to passenger {\it free speed}; for open ``lobby" spaces (i.e., the wider areas in the region of the main exit), this takes the value 1.542, for aisles it takes the value 1.0, and for seat rows we use the value 0.71 (all modifier values are based on \cite{gwynne2018small}). So, for example, a passenger with an allocated free speed of 2.5m/s may actually move at speeds up to 3.8m/s in lobby space. The occupant movement model used was {\it steering mode}{\footnote{\url{https://support.thunderheadeng.com/docs/pathfinder/2021-2/technical-reference-manual/}}, which allows for more realistic (and emergent) behaviours. All other parameters were left at the default settings in the Pathfinder software. The default behaviour profile simply requires agents to delay (in order to simulate the response/preparatory phase of noting the alarm and removing a seatbelt) and move to their assigned exit. The range of initial delay times was determined by combining ``seatbelt unfasten" and ``leave seat" timings taken from \cite{gwynne2018small} (mean 4.4s, s.d. 1.52s, min 2s, max 16s).

\begin{table}[]
\begin{centering}
\begin{tabular}{|l|l|l|}
\hline
\textbf{Parameter} & \textbf{Value} & \textbf{Note} \\
\hline
Free speed & 1.5-2.5m/s, uniform distribution & Observed:1.4-3.4m/s \cite{gwynne2018small} \\ \hline
Diameter & 45.58cm & See discussion in text \\ \hline
Initial orientation & 180 degrees & i.e., facing front \\ \hline
Initial delay & 2-16s, normal distribution & See discussion in text\\ \hline
\end{tabular}
\caption{{\it nobags} (default) passenger profile.}
\label{tab:nobags}
\end{centering}
\end{table}



\subsection{Additional baggage collection passenger behaviour}

In what follows, we use ``bag grab" as a shorthand term for the act of collecting and taking baggage during an emergency evacuation. For the specific baggage collection experiments, we introduce an additional passenger characteristic profile, {\it withbag}, which has exactly the same attribute set as the {\it nobags} profile, only the {\it Speed} range is reduced to a uniform distribution between 0.419-1.916 m/s, as specified in \cite{gwynne2018small}. The only major issue here is that we cannot {\it spatially} account for the items carried by passengers, as the software platform does not allow for coupled/carried objects. We attempted to address this by increasing the radius of the passengers deemed to be carrying baggage, but this caused jamming and other anomalous behaviour. We conclude that slowing relevant passengers down acts as a reasonable (if imperfect) proxy for baggage carrying.

In terms of baggage collection {\it behaviour}, we add an additional passenger behaviour, {\it getbag}. This is executed by all passengers who have the {\it withbag} profile, and simulates the effect of a passenger collecting their carry-on luggage from the overhead storage bins or from under their seat. The sequence of actions for this behaviour is as follows:

\begin{enumerate}
\item Implement initial delay (see above).
\item Go to nearest aisle or lobby region.
\item Wait for a randomly-selected period of time in order to simulate baggage collection, uniformly selected in the range 2.3-8.9s (as specified in \cite{gwynne2018small}).
\item Go to assigned exit.
\end{enumerate}

\subsection{Passenger distribution}

For each experiment, we average our results over 100 runs of the simulation. Passenger attributes are assigned  (according to the specified distributions) before every run, including the passengers selected to have the {\it withbag} profile (again, according to the specified bag grab probability). Where the level of occupancy is less than 100$\%$, the spatial distribution of passengers is also randomised before each run, by randomly removing passengers from a full configuration until the desired occupancy level is reached.

\section{Results}
\label{results}

In this Section we present the results{\footnote{Full datasets are available at \url{http://doi.org/10.17605/OSF.IO/2ZUKB}}} of two trials; the first (baseline) trial evaluates the performance of our model against the real-life certification test, and the second examines the impact of baggage collection behaviour on evacuation performance. Both trials used a full complement of 420 simulated passengers; for runs performed in the baggage collection trial, a specified proportion of passengers were randomly assigned the {\it withbag} profile.

In what follows, we use Total Evacuation Time (TET) as the main metric; this represents the time at which the last passenger clears an exit. For each experiment (i.e., examining a combination of specific occupancy level and bag grab proportion) within each trial, we ran the simulation 100 times, discarding as outliers any TETs that fell outside $2\times$ the standard deviation either side of the average evacuation time. This allowed us to discard anomalous runs in which jamming (or other unexpected behaviours) affected performance - in reality, this happened infrequently.  We generally set the X-axis (time) limit to a fixed value of 175 in order to allow for consistent comparison of curves. 

\subsection{Baseline trial results}

\begin{figure}[htb]
\begin{centering}
\includegraphics[width=\textwidth]{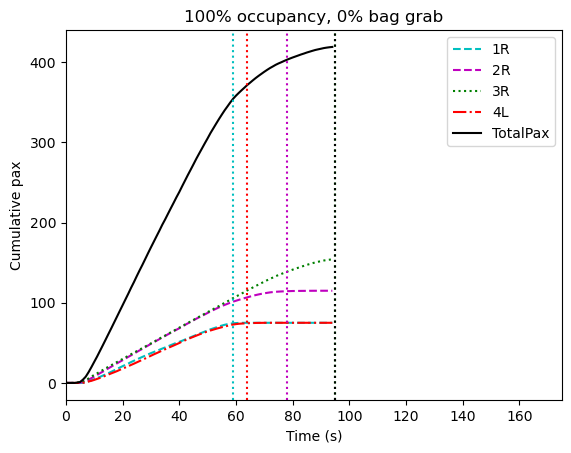}
\caption{Baseline trial: evacuation profiles for each exit (and total pax evacated) averaged over 100 runs. Dotted vertical lines denote average times at which each exit clears.}
\label{fig:all_exits_baseline}
\end{centering}
\end{figure}

For the baseline trial, we assume 100$\%$ seat occupancy and 0$\%$ ``bag grab" behaviour. In Figure ~\ref{fig:all_exits_baseline}, we show evacuation profiles (that is, the cumulative number of passengers, or {\it pax}, that have left the aircraft over time) for individual exits (plus the total), averaged over 100 runs. We denote with dotted vertical lines the average time at which each exit clears. 

For the baseline case, the mean TET was 92.95s (min 85.8s, max 93.6s, s.d 5.73), compared to the observed TET for the certification trial of 94.3s (or 85.7s, if we prefer to ignore the single outlier passenger). Whichever certification trial figure we choose, we believe that our TET is within acceptable tolerance.

We note that exits 1R and 4L have almost identical profiles, which is unsurprising, as both have 75 pax assigned to them.
The certification trial description \cite{Satterfield} reported that flow through exit 1R was interrupted 53.7s into the trial, for a period of 28s, but we did not observe this in our data. On average, exit 3R was the slowest to clear (this exit had the greatest number of assigned pax, at 155).

\begin{figure}[htb]
\begin{centering}
\includegraphics[width = 2.2in]{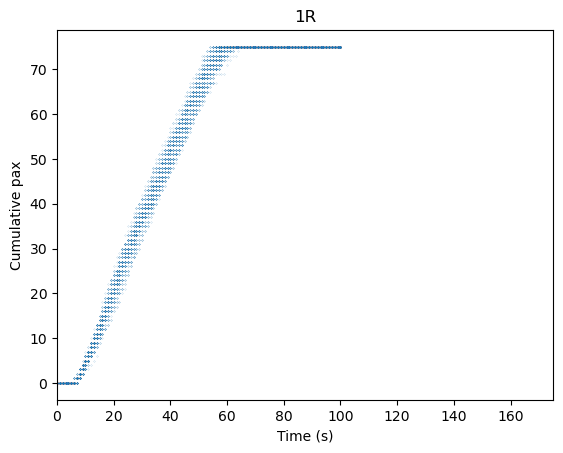}
\includegraphics[width = 2.2in]{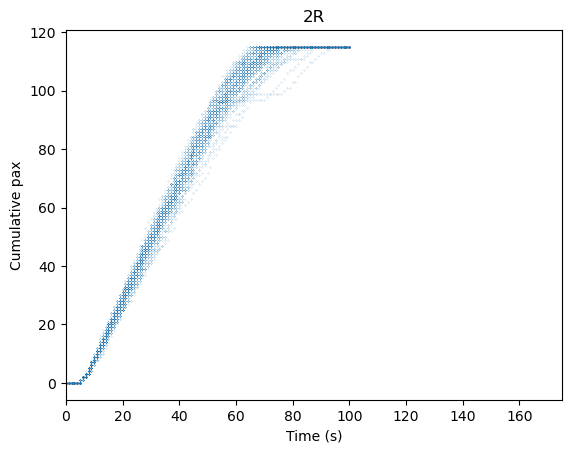}
\includegraphics[width = 2.2in]{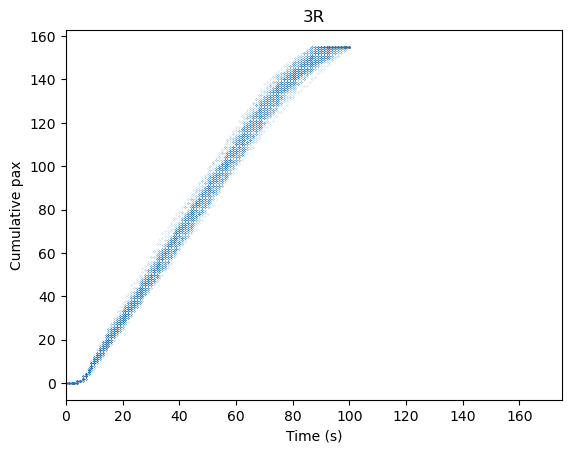}
\includegraphics[width = 2.2in]{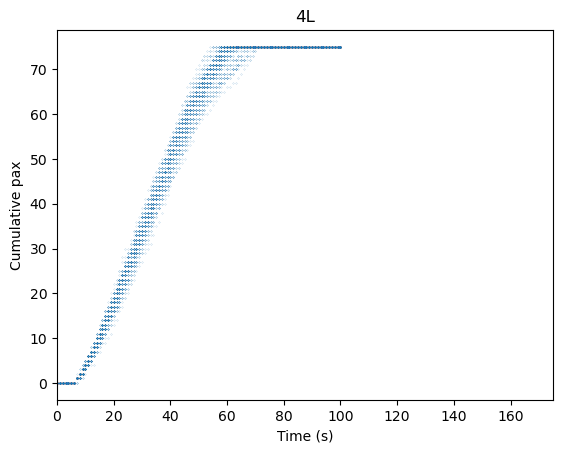}
\caption{Baseline trial: evacuation distributions for each exit over 100 runs.}
\label{fig:all_exits_variation}
\end{centering}
\end{figure}

In Figure ~\ref{fig:all_exits_variation}, we show the baseline case variation of flow over time for each exit. This shows the ``spread" of exit usage over time across 100 runs; we note that these bands are generally quite tight, with increased variation towards the end of the evacuation for exits 2R and 4L.

\subsection{Baggage collection trial results}

We first present the results of investigations into bag grab behaviour for 100$\%$ occupancy (i.e., as in the baseline case), looking at the impact of different bag grab probabilities. Again, for each value of bag grab (10-100$\%$, 10$\%$ increments) we ran the simulation 100 times. A summary of our results is supplied in Table ~\ref{tab:baggrab}. 

\begin{table}[]
\begin{centering}
\begin{tabular}{|l|l|l|l|l|}
\hline
{\bf \% bag grab} & {\bf Mean TET} & {\bf s.d} & {\bf Min TET} & {\bf Max TET} \\ \hline
10     & 98.26  &  5.37 & 88.5  & 112.6 \\ \hline
20     & 108.76 & 11.01 & 90.5  & 138.8 \\ \hline
30     & 121.08 & 14.72 & 96.0  & 171.1 \\ \hline
40     & 131.78 & 21.14 & 98.2  & 219.1 \\ \hline
50     & 136.02 & 14.99 & 107.9 & 173.4 \\ \hline
60     & 144.41 & 16.54 & 114.3 & 188.4 \\ \hline
70     & 153.19 & 18.68 & 122.2 & 197.6 \\ \hline
80     & 163.96 & 20.26 & 129.9 & 226.1 \\ \hline
90     & 176.28 & 25.64 & 137.7 & 270.2 \\ \hline
100    & 176.9  & 20.04 & 142.0 & 229.0 \\ \hline
\end{tabular}
\caption{Baggage collection trial: TET figures for different bag grab proportions (all 100$\%$ occupancy), averaged over 100 runs each.}
\label{tab:baggrab}
\end{centering}
\end{table}

In Figure ~\ref{fig:all_exits_half_baggrab} we show the evacuation time series for a representative 30$\%$ bag grab scenario. This shows that, using a reasonable approximation of bag grab frequency that has been observed in real-life incidents, only two exits (1R and 4L, the exits at the front and rear of the aircraft) clear before the ostensible ``safe" evacuation threshold of 90s.

\begin{figure}[ht!]
\begin{centering}
\includegraphics[width=3.5in]{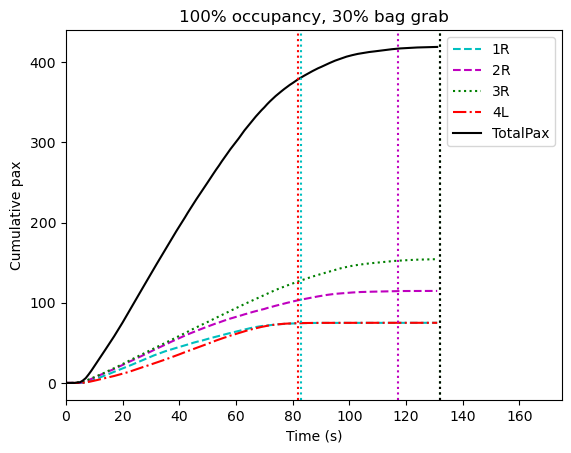}
\caption{Baggage collection trial: evacuation profiles for each exit for representative (30$\%$) bag grab case, averaged over 100 runs. Dotted vertical lines denote times at which each exit clears.}
\label{fig:all_exits_half_baggrab}
\end{centering}
\end{figure}

\begin{figure}[h!]
\begin{centering}
\includegraphics[width=3.5in]{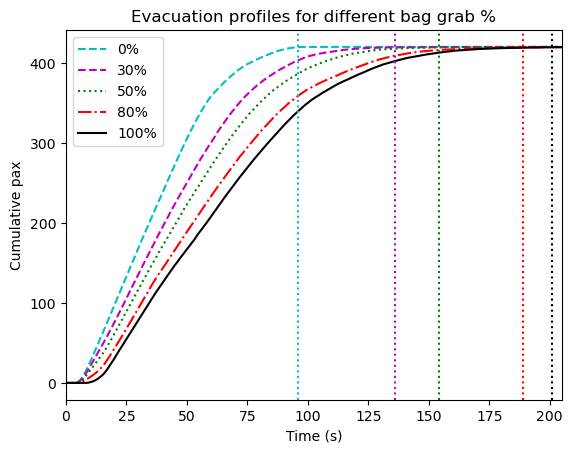}
\caption{Baggage collection trial: overall TET profiles for selected bag grab probabilities, averaged over 100 runs for each. Dotted vertical lines denote average times at which each exit clears.}
\label{fig:all_baggrab}
\end{centering}
\end{figure}

 \begin{figure}[h!]
\begin{centering}
\includegraphics[width=3.5in]{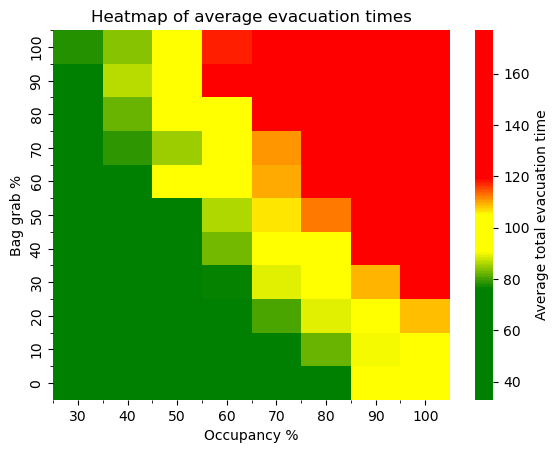}
\caption{Baggage collection trial: heatmap representation of TET for occupancy against bag grab.}
\label{fig:heatmap}
\end{centering}
\end{figure}

In Figure ~\ref{fig:all_baggrab} we show the time series for selected bag grab proportions (0$\%$, 30$\%$, 50$\%$, 80$\%$, 100$\%$).  For completeness, we also show a heatmap of TETS for all combinations of occupancy and bag grab probability (Figure ~\ref{fig:heatmap}). We observe an effective linear ``front" which indicates, for various levels of aircraft occupancy, the maximum baggage collection threshold that must not be passed if the average evacuation is to be conducted within the 90s time limit. The core message here is that for any aircraft occupancy level greater than 80$\%$ there exists {\it no} ``safe" level of baggage collection (if we define safety in terms of adherence to the 90s TET threshold), and the frequency of baggage collection that we observe in real-life events can cause potential safety issues for even relatively empty (60-70$\%$ full) flights. Although this seems intuitively obvious to observers, this is the first time that this behaviour has been rigorously quantified and characterised, especially in the context of wide-body aircraft.

\subsection{General observations}

The process of simulation construction and our results obtained highlighted some issues that are perhaps worthy of note. As previously mentioned, early iterations of the model increased the size of passengers carrying bags in order to make this more spatially realistic, as in, \cite{song2023emergency}, but this caused significant problems with jamming (in reality, a passenger carrying a bag will be able to manoeuvre it as a separate object, but the simulation platform treats the passenger plus their bag as a single mass, which is much less flexible). One possible improvement to future simulation platforms may allow for coupled objects in order to capture situations such as this. The problem of jamming was partially exacerbated by the platform's difficulty in handling constrained geometrical spaces, such as aisles. We note that a related study \cite{hrabak2022} using Pathfinder to model the evacuation of trains (i.e., a similarly constrained environment) presented similar challenges.

Two behaviours that were not considered in detail by our simulation were {\it explicit overtaking} and {\it aisle crossing}. Although agents were naturally allowed to move past one another if the environment allowed it, we did not factor in agents being able to ``squeeze past" one another in an aisle. We also initially considered allowing agents to cut across from one aisle to another in order to reduce their waiting time in a queue for an exit. However, in practice, this led to unrealistic behaviour, as agents in Pathfinder are unable to anticipate the problem of trying to pass an agent attempting to traverse the same narrow area in an opposite direction. This lead to  ``deadlock" behaviour, as two agents attempted to squeeze past ome another in opposite directions, with unrealistic persistence. 
 
\section{Conclusions and future work}
\label{conclusion}

In this paper we presented the first rigourous study of the impact of baggage collection behaviour on the evacuation of wide-body aircraft. We showed that even relatively low levels of baggage collection can significantly impact on evacuation time, especially in the case where an aircraft is full to capacity. We believe that this provides a firm foundation for future work.

In its 2020 report on the Lauda Air incident, the UK AAIB highlighted \cite{lauda} that ``The evidence from this accident, in combination with the collated evidence from previous cases shows that, even despite recent improvements, it remains the case that passenger briefing, safety cards and Flight Attendant instructions are insufficient to stop passengers retrieving cabin baggage during an evacuation. This hazard will still exist in future emergencies unless additional measures are taken to either reduce the impact of that behaviour on the safety and speed of an evacuation or to prevent passengers evacuating with baggage. Therefore, the following safety recommendation was made: that the European Union Aviation Safety Agency commission research to determine how to prevent passengers from obstructing aircraft
evacuations by retrieving carry-on baggage. The European Union Aviation Safety Agency has confirmed that the recommendation to commission
research into preventing passengers from obstructing aircraft evacuations by retrieving carry-on baggage will be considered for inclusion under the European Union Aviation Safety Agency Safety Risk Management process." 

One possible mitigation that has been discussed \cite{RAS2018} is the introduction of centralised locking of overhead bins, which would give cabin crew the ability to control passenger access during routine operations as well as emergency situations. Although this does offer one possible solution, it may also create unintended consequences, as frustrated passengers attempt to force open locked bins (it may also introduce additional complexity into aircraft infrastructure, with the possibility of future failure, and maintenance overheads). The current study offers a possible easy route to evaluating the impact, costs and benefits of such mitigations, as it would be straightforward to program into the simulation the relevant passenger behaviour(s).

More broadly, there has been much discussion over whether or not the 90s total evacuation time threshold for aircraft certification represents a realistic benchmark. At the time of writing, the Emergency Vacating of Aircraft Cabin Act has been presented to legislators in the US, with the aim of requiring the Federal Aviation Authority (FAA) to reconsider standards for evacuation, and to take into account ``real-life conditions" such as passenger anthropometry, passengers with reduced mobility and/or communication difficulties, seat configuration, pitch and width, and (crucially, in terms of the implications of this study) the presence of carry-on baggage. The debate over whether or not existing evacuation standards are fit for purpose is beyond the scope of the current paper, but we hope that our study will contribute to future discussion over a combined approach to more realistic standardisation, using computational studies that are validated with data obtained from ``real-life" trials.

\bibliography{evac}

\end{document}